\documentclass[10pt]{iopart}
\usepackage{iopams}
\usepackage{bm}
\begin{document}
\hbadness=10000

\title{Electromagnetic vortex lines riding atop null solutions of the Maxwell equations}
\author{Iwo Bialynicki-Birula}
\address{Center for Theoretical Physics, Polish Academy of Sciences,\\
Lotnikow 32/46, 02-668 Warsaw, Poland}
\ead{birula@cft.edu.pl}

\begin{abstract}
New method of introducing vortex lines of the electromagnetic field is
outlined. The vortex lines arise when a complex Riemann-Silberstein vector
$({\bm E} + \rmi{\bm B})/\sqrt{2}$ is multiplied by a complex scalar function
$\phi$. Such a multiplication may lead to new solutions of the Maxwell
equations only when the electromagnetic field is null, i.e. when both
relativistic invariants vanish. In general, zeroes of the $\phi$ function give
rise to electromagnetic vortices. The description of these vortices benefits
from the ideas of Penrose, Robinson and Trautman developed in general
relativity.

\end{abstract}

\submitto{\JOA}


\maketitle

\section{Introduction}

Complex scalar fields in a three-dimensional space carry generically a
collection of vortex lines. These vortex lines are located at the zeroes of the
field. Complex generic {\it vector fields} do not have such properties because
the requirement that all field components {\it simultaneously} vanish leads to
an overdetermined set of equations. There are at least two ways to overcome
this difficulty and to introduce vortex lines also for vector fields. One may
either select one relevant component of the field (as, for example, in the
scalar theory of light) or one may build a single field entity from various
vector components. The first approach has been used in most papers on phase
singularities in wave fields (for a thorough review and list of references see
\cite{den}). In our previous publication \cite{bb1} we have chosen the second
method. We have studied phase singularities and vortex lines associated with
the zeroes of the square ${\bm F}\!\cdot\!{\bm F}$ of the
Riemann-Silberstein\footnote{This name was introduced in a review paper
\cite{ibbwf} on the photon wave function, where one may also find information
on the history of the Riemann-Silberstein vector. The first to make a practical
use of the RS vector in the analysis of electromagnetic waves was Bateman
\cite{bat}. The RS vector offers a very convenient representation of the
electromagnetic field, especially in the study of vortex lines, and I shall
make an extensive use of it here.} (RS) vector (my units are chosen so that
$\epsilon = 1$ and $\mu = 1$)
\begin{eqnarray}
{\bm F} = ({\bm E} + \rmi{\bm B})/\sqrt{2}.\label{rs}
\end{eqnarray}
The vortex lines of the electromagnetic field defined in this manner are rather
elusive objects
---they lack clear signature and
may be hard to observe. Even though the phase of the field has dislocations,
the electromagnetic field {\em does not vanish} on these vortex lines, but it
is just a null electromagnetic field --- the two relativistic field invariants
vanish on vortex lines. Nevertheless, the vortex lines built on ${\bm
F}\!\cdot\!{\bm F}$ may still have some role to play in singular optics as
discussed in the papers by M. Berry \cite{berry} and G. Kaiser \cite{gerry}
appearing in this issue.

In the present paper I explore a different approach. It is based on the
observation that when a null solution of the Maxwell equations is taken as the
background field, an extra scalar multiplier may imprint on this solution a
rich vortex structure. This time {\it all components} of the electromagnetic
field will vanish on vortex lines. Of course, the vortex lines that are
introduced in this manner do not have a generic character since they are found
only in special cases --- for null fields. However, the class of solutions of
the Maxwell equations that carry these vortex lines is, in my opinion,
sufficiently broad and interesting to make this approach relevant for singular
optics. In addition, the present study touches upon some important concepts
discovered before in the general relativistic context.

\section{Scalar prefactor as a carrier of phase singularities}

The starting point of this investigation is an observation that a complex
scalar function $\phi({\bm r}, t)$ multiplying the RS vector will control the
zeroes of the electromagnetic field. Wherever $\phi({\bm r}, t)$ vanishes, the
electric and magnetic field vectors vanish. Assuming that the background field
${\bm F}$, as well as the product ${\phi\bm F}$, satisfy the Maxwell equations
\begin{eqnarray}
 \rmi\partial_t{\bm F}({\bm r}, t) = \nabla\times{\bm F}({\bm r},
 t),\;\;\;\nabla\!\cdot\!{\bm F}({\bm r}, t) = 0,\\
\label{max0}
 \fl \rmi\partial_t\left(\phi({\bm r}, t){\bm F}({\bm r}, t)\right) =
 \nabla\times\left(\phi({\bm r}, t){\bm F}({\bm r}, t)\right),
 \;\;\;\nabla\!\cdot\!\left(\phi({\bm r}, t){\bm F}({\bm r}, t)\right) = 0,
 \label{max}
\end{eqnarray}
we arrive at the following conditions on the prefactor field $\phi({\bm r}, t)$
(to simplify the notation, from now on I omit the dependence of ${\bm F}$ on
space-time coordinates)
\begin{eqnarray}
{\bm F}\,\rmi\partial_t\phi({\bm r}, t) = -{\bm F}\times\nabla\phi({\bm r},
t),\label{max1a}\\{\bm F}\!\cdot\!\nabla\phi({\bm r}, t) = 0.\label{max1b}
\end{eqnarray}
This set of equations possesses nontrivial solutions only when the background
field is null, i.e. ${\bm F}^2 = {\bm E}^2/2-{\bm B}^2/2+\rmi{\bm E}\cdot\!{\bm
B} = 0.$ In order to prove this assertion, we can take the scalar product of
both sides of Eq.~(\ref{max1a}) with the vector ${\bm F}$. Since the scalar
product on the right hand side is equal to zero, the left hand side must also
vanish and this results in the following alternative: either
$\partial_t\phi({\bm r}, t)$ or ${\bm F}\!\cdot\!{\bm F}$ must vanish. In the
first case, we end up with a trivial result: the scalar function $\phi({\bm r},
t)$ must be a constant. It cannot depend on space and time variables since its
gradient is at the same time parallel to ${\bm F}$, as seen from
Eq.~(\ref{max1a}), and orthogonal to ${\bm F}$, as seen from Eq.~(\ref{max1b}).
In the second case we may have a nontrivial form of $\phi({\bm r}, t)$. Taking
the scalar product of Eq.~(\ref{max1a}) with the complex conjugate RS vector
${\bm F}^*$ we obtain the condition
\begin{eqnarray}
\fl \rmi{\bm F}^*\!\cdot\!{\bm F}\;\partial_t\phi({\bm r}, t) = -{\bm
F}^*\!\cdot\!\left({\bm F}\times\nabla\phi({\bm r}, t)\right) = -\left({\bm
F}^*\times{\bm F}\right)\!\cdot\!\nabla\phi({\bm r}, t).\label{max21}
\end{eqnarray}
Finally, I obtain the following two partial differential equations for the
function $\phi({\bm r}, t)$
\begin{eqnarray}
\partial_t\phi({\bm r}, t) + {\bm n}\!\cdot\!\nabla\phi({\bm r}, t) = 0,
\;\;\;{\bm F}\!\cdot\!\nabla\phi({\bm r}, t) = 0,\label{max2}
\end{eqnarray}
where ${\bm n}$ is the normalized Poynting vector
\begin{eqnarray}
{\bm n} = \frac{-\rmi{\bm F}^*\times{\bm F}}{{\bm F}^*\!\cdot\!{\bm F}} =
\frac{{\bm E}\times{\bm B}}{{\bm E}^2/2+{\bm B}^2/2}.\label{unit}
\end{eqnarray}
The solutions of Eqs.~(\ref{max2}) will in general have zeroes and this will
lead to vortex lines of the electromagnetic field riding atop the background
solution ${\bm F}$. On each vortex line the electromagnetic field vanishes.
Near the vortex line the electric and magnetic field vectors followed around a
closed contour rotate by $2\pi m$, where $m$ is the topological charge of the
vortex. In contrast to nonrelativistic wave mechanics \cite{bbs,bmrs}, there is
no interaction between vortex lines introduced in this manner. All vortices
move independently, since the product of solutions of the first order partial
differential equations (\ref{max2}) is again a solution. We may always add new
vortices without changing the motion of the existing ones. In the following
sections I shall give general solutions of Eqs.~(\ref{max2}) in two special
cases: when the background field is a plane monochromatic wave and when it is
the Robinson-Trautman field.

\section{Simple example}

Let us consider the solution of the Maxwell equations described by the
following RS vector
\begin{eqnarray}
{\bm F}({\bm r}, t) = (\hat{\bm x}+\rmi\hat{\bm y})\exp(\rmi kz - \rmi \omega
t).\label{pwsoln}
\end{eqnarray}
This solution describes the left-handed circularly polarized wave propagating
in the $z$ direction. It is a null field, since $(\hat{\bm x}+\rmi\hat{\bm
y})\!\cdot\!(\hat{\bm x}+\rmi\hat{\bm y})=0$. In this case, Eqs.~(\ref{max2})
take on the form
\begin{eqnarray}
(\partial_t + \partial_z)\phi({\bm r}, t) = 0,\;\;\; (\partial_x +
\rmi\partial_y)\phi({\bm r}, t) = 0.\label{max2a}
\end{eqnarray}
General solution of these equations is an arbitrary function of two variables:
the real variable $z-t$ and the complex variable $x+\rmi y$. Thus, we arrive at
the family of solutions of the Maxwell equations of the form
\begin{eqnarray}
{\bm F}({\bm r}, t) = f(z-t,x+\rmi y)(\hat{\bm x}+\rmi\hat{\bm y})\exp(\rmi k(z
- t)).\label{fam1}
\end{eqnarray}
These solutions may have a rich vortex structure. Assuming, for simplicity, a
polynomial dependence on $x+\rmi y\equiv w$, we may write down the function $f$
in the factorized form
\begin{eqnarray}
\fl f(z-t,x+\rmi y) = (w-a_1(z-t))(w-a_2(z-t))\dots(w-a_n(z-t)),\label{factor}
\end{eqnarray}
where $a_k(z-t)$ are some functions of $z-t$. Each term in this product gives
rise to an infinite vortex line whose shape is determined by the equation
$x+\rmi y=a_k(z-t)$. Each vortex line moves independently. In general, each
vortex line flies with the speed of light in the $z$-direction. However, when
the function $a_k(z-t)$ is just a constant, the corresponding vortex line is a
stationary straight line in $z$-direction. A simple example of the $\phi$
function of the type (\ref{factor}) is
\begin{eqnarray}
 \phi({\bm r}, t) = (x+\rmi y)^m.\label{example1}
\end{eqnarray}
In this case the RS vector $\phi{\bm F}$ carries a straight vortex line along
the $z$-axis with the topological charge $m$.

\section{Trautman-Robinson fields}

Null solutions of the Maxwell equations were extensively studied by Robinson
\cite{rob} in connection with his work in general relativity. In particular,
these solutions played an important role in the discovery of new solutions of
Einstein field equations \cite{robtr} and also they appear in Penrose twistor
theory \cite{penrin}. A family of null solutions of the Maxwell equations that
featured prominently in these studies was described in \cite{robtr1}
--- I shall call them the Robinson-Trautman (RT)
fields\footnote{Topical review has been recently published by Trautman
\cite{tr1}}. They are usually expressed in a special coordinate system but I
shall use here more intuitive Cartesian coordinates. In these coordinates the
RS vectors representing the RT solutions can be written in the form
\numparts
\begin{eqnarray}
 F_x = f(\alpha,\beta)\frac{\beta^2-1}{t-\rmi a-z},\\
 F_y = \rmi f(\alpha,\beta)\frac{\beta^2+1}{t-\rmi a-z},\\
 F_z = -f(\alpha,\beta)\frac{2\beta}{t-\rmi a-z},\label{rt}
\end{eqnarray}
\endnumparts
where $f(\alpha,\beta)$ is an arbitrary function of the following complex
combinations of the Cartesian coordinates
\begin{eqnarray}
 \alpha=t-\rmi a+z-\frac{x^2+y^2}{t-\rmi a-z},\;\;\beta=\frac{x-\rmi y}{t-\rmi a-z}.
\end{eqnarray}
The RT Maxwell fields are related to the simple plane wave field discussed in
the previous section. Taking the $f$ function in the form
$f(\alpha,\beta)=\alpha^{-3}$, we obtain the following localized null solution
of Maxwell equations
\numparts
\begin{eqnarray}
 F_x = \left((x-\rmi y)^2 - (t-\rmi a -z)^2\right)/d^3,\label{ibb1}\\
 F_y = \rmi\left((x-\rmi y)^2 + (t-\rmi a -z)^2\right)/d^3,\label{ibb2}\\
 F_z = -2(x-\rmi y)(t-\rmi a -z)/d^3,\label{ibb3}
\end{eqnarray}
\endnumparts
where $d=((t-\rmi a)^2 - x^2 - y^2 - z^2)$. This field may be obtained from the
plane wave solution of the previous section by the coordinate transformation
(conformal reflection) $x^{\mu}\to x^{\mu}/x^2$ accompanied by the shift in
time by an imaginary constant $-\rmi a$ and evaluated in the limit of infinite
wavelength, when $k \to 0$. The energy density of this field is equal to
\begin{eqnarray}
 {\bm F}^*\!\cdot\!{\bm F} = 2\frac{\left(a^2+(t-z)^2+x^2+y^2\right)^2}
 {\left((a^2-t^2+x^2+y^2+z^2)^2+4a^2t^2\right)^3}.\label{energy}
\end{eqnarray}
From this formula we see that the electromagnetic field
(\ref{ibb1}--\ref{ibb3}) can never vanish. Note, that the presence of $\rmi a$
eliminates the singularity on the light cone and makes the energy of the field
finite,
\begin{eqnarray}
 \int\!d^3r{\bm F}^*\!\cdot\!{\bm F} = \frac{\pi}{4a^5}.\label{totenergy}
\end{eqnarray}
There are some simple solutions with a single vortex line built on this
background function. The simplest ones are obtained by taking the scalar
multiplier function $\phi$ equal to $\beta^m$. These solutions have a
stationary vortex line along the $z$-direction with the topological charge
equal to $-m$. A bit more complex vortex line is obtained when $\phi =
\alpha+4a\beta$. The vanishing of $\phi$ leads to the following two equations
for the vortex coordinates
\begin{eqnarray}
 2y = t,\;\;\;(x+2a)^2+z^2 = 3a^2+\frac{3}{4}t^2.
\end{eqnarray}
Thus, at each time the vortex line forms a circle lying on the uniformly moving
$y=t/2$ plane. The radius of the circle is shrinking until $t=0$ and then it
starts expanding. Possibilities for constructing solutions with more intricate
vortex lines are unlimited.

\ack{I would like to thank Michael Berry, Mark Dennis and Gerald Kaiser for
discussions and fruitful comments. I am especially indebted to Andrzej Trautman
who shared with me his profound understanding of complex spacetime.}


\section*{References}
\begin{thebibliography}{22}
\bibitem{den} Dennis M R 2001 {\it Topological Singularities in Wave Fields}
Ph.D. Thesis Physics University of Bristol
(http://rogers.phy.bris.ac.uk/mrdthesis.pdf)
\bibitem{bb1} Bialynicki-Birula I and Bialynicka-Birula Z 2003 \PR A {\bf 67} 062114
\bibitem{ibbwf} Bialynicki-Birula I 1996 in {\it Progress in Optics},
 Vol. XXXVI edited by E. Wolf (Amsterdam: Elsevier)
\bibitem{bat} Bateman H 1955 {\it The Mathematical Analysis of Electrical and Optical
Wave-Motion on the Basis of Maxwell's Equations} (New York: Dover)
\bibitem{berry} Berry M V J. Opt. A: Pure Appl. Opt. (this issue)
\bibitem{gerry} Kaiser G J. Opt. A: Pure Appl. Opt. (this issue)
\bibitem{bbs} Bialynicki-Birula I, Bialynicka-Birula Z and {\'S}liwa C 2000
Phys. Rev. A {\bf 61} 032110
\bibitem{bmrs} Bialynicki-Birula I, M{\l}oduchowski T, Radozycki T
and {\'S}liwa C 2001 Acta Phys. Pol. A {\bf 100} (Supplement) 29
\bibitem{rob} Robinson I 1961 J. Math. Phys. {\bf 2} 290
\bibitem{robtr} Robinson I and Trautman A 1960 Phys. Rev. Lett. {\bf 4} 431; 1962 Proc.
Roy. Soc. {\bf 265} 463
\bibitem{penrin} Penrose R and Rindler W 1986 {\it Spinors and Space-Time} Vol
2 (Cambridge: Cambridge University Press)
\bibitem{robtr1} Robinson I and Trautman A 1989 in {\it New Theories in Physics:
Proc. Warsaw. on Elementary Particle Physics} ed Z Ajduk {\it et al}
(Singapore: World Scientific) pp 454-97
\bibitem{tr1} Trautman A 2002 Class. Quantum. Grav. {\bf 19} R1

\end{thebibliography}
\end{document}